\documentclass[reprint,prl,twocolumn,showpacs,showkeys,floatfix,superscriptaddress,preprintnumbers]{revtex4} 
\usepackage{graphicx}                              
\usepackage{amsmath,amsfonts}
\graphicspath{{./pictures/}}                       
 
\usepackage{dsfont}   
\usepackage{epsfig}
\usepackage{amssymb}
\usepackage{amsfonts}
\usepackage{float}
\usepackage{bbm}
\usepackage{psfrag}
\usepackage{bbold}

\newcommand{\bdm}{\begin{displaymath}}
\newcommand{\edm}{\end{displaymath}}
\newcommand{\beq}{\begin{equation}}
\newcommand{\eeq}{\end{equation}}
\newcommand{\bea}{\begin{eqnarray}}
\newcommand{\eea}{\end{eqnarray}}
\newcommand{\bit}{\begin{itemize}}
\newcommand{\eit}{\end{itemize}}
\newcommand{\bc}{\begin{center}}
\newcommand{\ec}{\end{center}}
\newcommand{\re}{\relax{\rm I\kern-.18em R}}
\newcommand{\ID}{\mathbbm{1}}

\newcommand{\Dov}{{\cal D}^{(ov)}}
\newcommand{\D}{\Dov}

\newcommand{\fermiMat}{{\cal M}}

\newcommand{\latVol}[2]{\ifnum #1=#2 $#1^4$ \else $#1^3\times #2$\fi}
\newcommand{\lattice}[2]{\ifnum #1=#2 $#1^4$-lattice \else $#1^3\times #2$-lattice\fi}
\newcommand{\latticeX}[3]{\ifnum #1=#2 $#1^4$-lattice#3 \else $#1^3\times #2$-lattice#3\fi}
\newcommand{\lattices}[2]{\ifnum #1=#2 $#1^4$-lattice \else $#1^3\times #2$-lattices\fi}

\newcommand{\fig}[1]{Fig.~\ref{#1}}

\newcommand{\Ref}[1]{Ref.~\cite{#1}}

\newcommand{\Tr}{\mbox{Tr}}
\newcommand{\RE}{\mbox{Re}}

\newcommand{\SUtwoTimesUoneY}{{SU(2)_\mathrm{L}\times U(1)_\mathrm{Y}}}

\newcommand{\dslash}{\ensuremath\partial\kern-0.53em/}

\begin{document}
\preprint{DESY 13-006}
\preprint{CERN-PH-TH/2013-007}
\preprint{HU-EP-13/1}

\title{Constraining a fourth generation of quarks: \\
non-perturbative Higgs boson mass bounds} 

\author{J. Bulava}\email{john.bulava@cern.ch}
\affiliation{CERN, Physics Department, 1211 Geneva 23, Switzerland}
\author{K. Jansen}\email{karl.jansen@desy.de}
\affiliation{NIC, DESY,
 Platanenallee 6, D-15738 Zeuthen, Germany}
\author{A. Nagy}\email{attila.nagy@physik.hu-berlin.de}
\affiliation{NIC, DESY,
 Platanenallee 6, D-15738 Zeuthen, Germany}
\affiliation{Humboldt-Universit\"at zu Berlin, Institut f\"ur Physik,
Newtonstr. 15, D-12489 Berlin, Germany}

\date{\today}

\begin{abstract}
We present a non-perturbative determination of the upper and lower  
Higgs boson mass bounds with a heavy fourth 
generation of quarks from numerical lattice computations in a 
chirally symmetric Higgs-Yukawa model. We find that the upper bound only 
moderately rises with 
the quark mass while the lower bound increases significantly,
providing additional constraints on the existence of a 
straight-forward fourth quark generation. 
We examine the stability of the lower bound under the  
addition of a higher dimensional operator to the scalar field potential using perturbation theory,  
demonstrating that it is not significantly altered for small values of the coupling of this operator. For a Higgs boson 
mass of $\sim125\mathrm{GeV}$ we find that the
maximum value 
of the fourth generation quark mass  
is $\sim300\mathrm{GeV}$, which is already 
in conflict with bounds from direct searches. 

\end{abstract}


 
\maketitle

\section{Introduction}
\label{sec:Introduction}
 
A heavy fourth generation of quarks is an attractive and simple 
extension (denoted SM4) of the three-generation Standard Model 
(SM3)~\cite{Holdom:2009rf,Carena:2004ha}. However, heavy fermion 
effects are expected to significantly contribute to the properties of the 
Higgs boson, leading to measurable deviations in Higgs production cross 
sections and branching ratios. The recent discovery of a scalar boson 
seemingly consistent with the Standard Model expectation therefore casts serious doubt on 
straight-forward 
fourth generation scenarios~\cite{Eberhardt:2012ck}.  
 
Another property of the Higgs boson which is sensitive to a possible fourth 
generation is its mass. Invoking arguments from 
perturbation theory, the Higgs boson mass is bounded from above by the 
triviality bound, which reflects the Gaussian nature of the UV 
fixed point, and from below by the vacuum instability bound, which ensures that 
the theory has a stable vacuum state. 

In perturbation theory, the lower bound can be obtained by examining the 
effective potential and demanding that it remains bounded from below. As the 
fermion fields contribute negatively to the effective potential, 
they have a destabilizing effect which leads (by demanding the stability 
of the theory) to a lower 
Higgs boson mass bound. However it is expected that the 
perturbative expansion breaks down for Yukawa couplings near or less than  
the perturbative unitarity bound~\cite{Denner:2011vt}, which allows maximal fermion
masses of roughly \newline 
$m_f \sim 500-600\mathrm{GeV}$~\cite{Chanowitz:1978uj,Chanowitz:1978mv}.   

Due to these considerations, study of heavy fourth generation extensions to the Standard Model necessitates non-perturbative methods.  
To this end, we employ lattice field theory techniques which allow us to 
compute lower and upper Higgs boson mass bounds as well as resonance parameters 
of the Higgs boson non-perturbatively. Such a strategy has already 
been applied in Refs.~\cite{Gerhold:2009ub,Gerhold:2010bh,Gerhold:2010wy,Gerhold:2011mx} 
for the non-perturbative determination of the upper and lower Higgs boson mass bounds and 
the Higgs boson resonance parameters in the SM3. 
First results for a fourth quark generation have been presented in Ref.~\cite{Gerhold:2010wv}. 
Investigations of a bulk-phase transition at very large values 
of the Yukawa coupling have been initiated in
Ref.~\cite{Bulava:2011jp} while first studies of the system at non-zero temperatures 
can be found in  
Ref.~\cite{Bulava:2011ss}. 
For a summary of recent work on this model see Ref.~\cite{Bulava:2012rb}.

The lattice calculations are possible due to 
a lattice discretization of the fermion action which respects an exact chiral symmetry at finite lattice spacing $a$~\cite{Luscher:1998pqa}, 
thus ensuring that the chiral 
character of the Higgs-fermion couplings is respected 
by the lattice regulator
in a conceptually clean manner. 
This advance has 
triggered a number of lattice investigations of Higgs-Yukawa like 
models
\cite{Bhattacharya:2006dc,Giedt:2007qg,Poppitz:2007tu,Gerhold:2007yb,Gerhold:2007gx,Fodor:2007fn,Gerhold:2009ub,Gerhold:2010wy,Gerhold:2010bh}.

In this letter we report on a systematic investigation of the lower and 
upper Higgs boson mass bounds for quark masses ranging from the Standard 
Model top quark mass of $m_f = m_t\approx 175 \mathrm{GeV}$ up to masses of 
$m_f\approx 700\mathrm{GeV}$, which is above the perturbative unitarity bound. Our 
calculations are performed at a fixed lattice cutoff 
of $\Lambda = \frac{1}{a} \approx 1.5 \mathrm{TeV}$  
which ensures that the fermion and Higgs boson masses are sufficiently 
far from the cutoff scale. 
 We also rely heavily on the techniques and simulation strategies 
of Refs.~\cite{Gerhold:2009ub,Gerhold:2010bh,Gerhold:2010wy,Gerhold:2011mx}. 

Our results indicate that with increasing quark mass there is a substantial 
upward shift of the lower Higgs boson mass bound leading to severe
constraints on a straight-forward fourth quark generation in light of the 
recently discovered scalar boson when interpreted as the 
Standard Model Higgs boson. 
In order to examine the stability of our results, we analyze the 
effect of 
including a higher dimensional operator in the scalar field potential using 
perturbation theory. For small values of the coupling of this term we 
confirm that the lower bound is not significantly altered. 

\section{The Model}
\label{sec:model}

Here we briefly review the definition of our model. For a more complete 
treatment, see Ref.~\cite{Gerhold:2010wy}. 
We employ a lattice discretization  
 of the Standard Model Higgs-fermion sector which exactly respects a chiral symmetry at finite lattice spacing.  
As the dynamics of the complex scalar (Higgs) doublet are expected to be 
dominated by its interactions with the heaviest fermions, we consider a single degenerate quark
doublet only.
 By the same reasoning, as well as for computational simplicity,  we also 
neglect gauge fields. 

One of the defining features of the Standard Model is the chiral 
structure of the scalar-fermion interactions. In order to reproduce this 
structure in a lattice discretization it is crucial to maintain a chiral
symmetry at finite lattice spacing. 
 This has been a long-standing obstacle to the lattice regularization of Higgs-Yukawa models and was finally overcome 
 by employing the Neuberger 
 `Overlap'~\cite{Luscher:1998pqa, Neuberger:1997fp,Neuberger:1998wv} 
 discretization of the fermion action. 

Following the proposition in Ref.~\cite{Luscher:1998pqa} we can therefore construct a 
lattice Higgs-Yukawa model with a global 
$SU(2)_L \times U(1)_Y$ symmetry.
Specifically, the fields included in our model are a scalar 
doublet $\varphi$ and two fermions, the left-handed components of 
which are paired into an $SU(2)$ doublet. 
The lattice action can
thus be written as
\begin{align}
\label{eq:DefYukawaCouplingTerm}
S &= S_F + S_{\Phi}, \quad S_F =
\sum_{xy} \bar\psi_{x}\, \fermiMat_{xy}\, \psi_{y} ,\\ 
S_\Phi  &=  -\kappa\sum_{x,\mu} \Phi_x^{T} \left[\Phi_{x+\mu} + \Phi_{x-\mu}\right]
+ \sum_{x} \Phi^{T}_x\Phi_x \nonumber \\\nonumber
& + \hat\lambda \sum_{x} \left(\Phi^{T}_x\Phi_x - 1 \right)^2,\\\nonumber
\fermiMat_{xy}  &= \D_{xy}\mathbb{1}_{2\times2}  +    
y\big(P_+ \phi^\dagger_x \hat P_+ + P_- \phi_x \hat P_-\big)\delta_{xy},
\end{align}
where $\psi$ is a doublet of four-component spinor fields, $\D$ is the free 
Overlap Dirac operator with a Wilson 
kernel, $\Phi^{\mu}_x \in \re^4$, and $\phi_x = \Phi^{\mu}_x \tau_\mu$, with 
$\tau_{\mu} = (1,-i\vec{\sigma})$. The left- and right-handed projection 
operators $P_{\pm}$ and the 
modified projectors $\hat P_{\pm}$ are given by
\begin{align}\label{eq:proj}
&P_\pm = \frac{1 \pm \gamma_5}{2}, \quad
\hat P_\pm = \frac{1 \pm \hat \gamma_5}{2}, 
\\\nonumber
&\hat\gamma_5 = \gamma_5 \left(\ID - \frac{1}{\rho} \D \right).
\end{align}

The action introduced above obeys an exact global 
$\mbox{SU}(2)_L\times U(1)_Y$ 
lattice chiral symmetry. For  
$\Omega_L\in \mbox{SU}(2)$ and $\theta \in [0, 2\pi]$ the action is invariant under the transformation
\bea
\label{eq:ChiralSymmetryTrafo1}
\psi & \rightarrow  &  U_Y \hat P_+ \psi + U_Y\Omega_L \hat P_- \psi,
 \nonumber \\
\bar\psi & \rightarrow &  \bar\psi P_+ \Omega_L^\dagger U_{Y}^\dagger + \bar\psi P_- U^\dagger_{Y}, \nonumber \\
\phi & \rightarrow &  U_Y  \phi \Omega_L^\dagger,
\phi^\dagger \rightarrow \Omega_L \phi^\dagger U_Y^\dagger
\eea
with $U_{Y} \equiv \exp(i\theta Y)$, where $Y$ labels the representation of the 
global hypercharge symmetry group $U(1)_Y$. It should be noted that in the continuum 
limit the (global) 
continuum $\mbox{SU}(2)_L\times \mbox{U}(1)_Y$ chiral symmetry is recovered.

This formulation enables a numerical
study of the limit $\hat{\lambda} = \infty$ on the lattice by simply 
enforcing the constraint $\Phi^{T}_x\Phi_x = 1, \:  \forall x$. 
Also, we can relate the parameters and fields appearing in 
Eq.~\ref{eq:DefYukawaCouplingTerm} 
to those appearing in the standard scalar complex doublet continuum Lagrangian 
($\mathcal{L} = |\partial_{\mu} \varphi |^2 + \frac{1}{2}m^2_0|\varphi|^2 + \lambda|\varphi|^4$) by  
\beq \nonumber
\label{eq:RelationBetweenHiggsActions}
\varphi_x = \sqrt{2\kappa}
\left(
\begin{array}{*{1}{c}}
\Phi_x^2 + i\Phi_x^1\\
\Phi_x^0-i\Phi_x^3\\ 
\end{array}
\right), 
\eeq
\beq \nonumber
\lambda=\frac{\hat\lambda}{4\kappa^2}, \qquad
m_0^2 = \frac{1 - 2\hat\lambda-8\kappa}{\kappa}.
\eeq

\section{Computational Strategy}
\label{sec:simstrag}

The cutoff in our model is 
provided by 
the inverse lattice spacing ($\Lambda=1/a$) and we set the physical value 
of $\Lambda$ 
using the phemonenological Higgs field vacuum expectation value 
($v_R$), 
\bea
\label{eq:FixationOfPhysScale}
(\sqrt{2} G_F )^{-\frac{1}{2}} \sim   
246\, \mbox{GeV} &=& \frac{v_R}{a} \equiv \frac{v}{\sqrt{Z_G}\cdot a},
\eea
where $Z_G$ denotes the Goldstone boson field renormalization constant and $v$ 
the bare scalar field vacuum expectation value.
We target a mass range for the degenerate fermion doublet of  
$175\mathrm{GeV} \lesssim m_f \lesssim 700\mathrm{GeV}$, while fixing 
the cutoff at $\Lambda \approx 1.5\mathrm{TeV}$. 
Although the masses of the fermion doublet are degenerate in this work, we plan 
to assess the effect of a mass splitting in the near future. At $m_f = m_t = 175\mathrm{GeV}$ the splitting $m_b-m_t$ has been taken into account and found to have a small effect on the lower Higgs boson mass 
bound which, moreover, can be taken into account by the effective
potential evaluated in lattice perturbation theory~\cite{Gerhold:2009ub}.  

We now briefly discuss the simulation algorithm. More details can be found in 
Ref.~\cite{Gerhold:2010wy}. 
The Monte-Carlo simulations are carried out using a variant~\cite{Frezzotti:1997ym} of the 
Hybrid Monte Carlo (HMC) algorithm~\cite{Duane:1987de} in which the integration over the 
Grassman fields is done analytically at the expense of introducing the 
determinant of the fermion action in the updating. Therefore, our remaining degrees of freedom in the Monte Carlo integration are the scalar fields. 

Due to the absence of gauge 
fields, the Overlap operator of Eq.~(\ref{eq:DefYukawaCouplingTerm}) can be 
constructed exactly in momentum space. 
However as the Yukawa coupling is local in position space, Fast-Fourier-Transforms (FFT's) are employed to efficiently apply this term. 
Finally, Fourier acceleration~\cite{Davies:1987vs} is used to propagate the 
low momentum modes using coarser time-steps along the HMC evolution, which effectively reduces 
the autocorrelation between successive scalar field configurations.      

It has been demonstrated~\cite{Gerhold:2009ub} that the Higgs boson mass is
a monotonically increasing function of the bare quartic coupling 
$\hat{\lambda}$. Therefore, the lower bound for the Higgs boson mass at fixed
cutoff and $m_f$ is obtained at $\hat{\lambda} = 0$, while the upper bound is 
obtained at $\hat{\lambda} = \infty$~\cite{Gerhold:2010bh}. 

Since we work in a finite volume with no external symmetry breaking source, the naively defined vacuum expectation value is zero in an ensemble average. 
We therefore resort to a special 
technique, pioneered in Ref.~\cite{Fukuda:1974ey} and employed in 
Refs.~\cite{Hasenfratz:1989ux,Hasenfratz:1990fu,Gockeler:1991ty}, 
of rotating a given scalar field configuration to a preferred direction. It 
can be shown that in infinite volume this leads 
to the same vacuum expectation value as obtained in the standard procedure 
involving the limit of vanishing external source~\cite{Gockeler:1991ty}.  

The determination of the Goldstone boson renormalization constant $Z_G$, the Higgs boson mass 
$m_H$, and the fermion mass $m_{f}$ have been discussed in 
detail in \Ref{Gerhold:2010bh} and are briefly reviewed here. 
The renormalization constant $Z_G$ is computed from the slope of the inverse 
Goldstone boson propagator at 
vanishing Euclidean four-momentum transfer, in the standard on-shell scheme. 
Operationally, this constant is obtained from fits to the propagator at 
small momenta.   

Due to the existence of massless Goldstone modes in the spontaneously broken 
phase of our theory, the Higgs boson is unstable and can decay to final states 
containing an even number of Goldstone bosons.  
We employ two definitions of the Higgs boson mass which both ignore this finite 
decay width. The Higgs boson propagator mass $m^{P}_H$ is derived from fits to the momentum space Higgs propagator defined as
\begin{equation}
\tilde G_H(p) = \langle \tilde h_p \tilde h_{-p}\rangle, 
\;\; \tilde h_p = \frac{1}{\sqrt{L_s^3\cdot L_t}}\sum\limits_x e^{-ipx} h_x \nonumber 
\end{equation}
while the correlator mass $m^{C}_H$ is derived from fits to the Higgs temporal 
correlation function. Although finite decay width corrections to these formulae are quite different, the mass extracted from both of these procedures typically differs by less than 10\%, lending credence to our 
approximation of a stable Higgs boson. Furthermore, a rigorous study of the 
Higgs boson decay width at non-vanishing external source has been performed at 
$m_f = m_t \sim 175\mathrm{GeV}$ in Ref.~\cite{Gerhold:2011mx} which obtained 
a narrow decay width for all values of the bare quartic coupling, further 
supporting the validity of the stable Higgs boson approximation. 
For this work we quote $m^{P}_{H}$ as the central value as it is typically the 
lowest estimate of $m_H$.  

Finally, we compute the quark mass from the exponential 
decay of the temporal correlation function $C_f(t)$ at large Euclidean time 
separations $t$, defined as 
\begin{equation}
\label{eq:DefOfFermionTimeSliceCorr}
C_f(t) = \frac{1}{L_s^6} \sum\limits_{\vec x, \vec y}
\,\RE\,\Tr\,\left(f_{L,0, \vec x}\cdot \bar f_{R,t,\vec y}\right),
\end{equation}
where the left- and right-handed spinors are defined using the projection 
operators in Eq.~\ref{eq:proj}.

\section{Numerical Results}
\label{sec:numericalresults}

As mentioned above, massless Goldstone excitations are present in the 
spontaneously broken phase of our model. Since we work at zero external symmetry-breaking source,  
finite size effects are not exponentially suppressed but rather algebraic in nature and  
 proportional to inverse even powers of the linear extent of the lattice, i.e. 
 $\mathcal{O}(1/L_s^2)$ at leading order~\cite{Hasenfratz:1989ux,Hasenfratz:1990fu,Gockeler:1991ty}. 
These finite size effects can be quite substantial  
and an infinite volume extrapolation of our results 
for the renormalized vacuum expectation value and all masses is required. 

\begin{figure}[htb]
\centering
\includegraphics[width=0.40\textwidth]{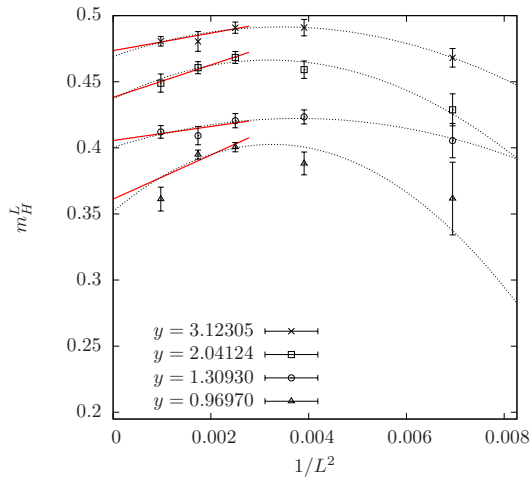}
\caption{Infinite volume extrapolation of the Higgs boson mass extracted from fits 
to the momentum space Higgs propagator ($m^{P}_{H}$). The data  
are for the Higgs boson mass relevant for 
the upper bound ($\hat{\lambda} = \infty$) and a  
range of 
bare Yukawa couplings corresponding to fermion masses from the physical top quark mass to $\sim 700\mathrm{GeV}$.  
The infinite volume 
extrapolation is performed by fitting the data to a 
linear function in $1/L_s^2$ 
taking only data with 
$L_s\ge 16$ into account.
\label{fig:FiniteSizeEffects}
}
\end{figure}

Finite volume data for the Higgs boson propagator mass ($m^{P}_H$) are shown in \fig{fig:FiniteSizeEffects} as 
an example of our infinite volume extrapolations. The lattice data are plotted 
versus $1/L_s^2$ and 
extrapolated to the infinite volume limit by means of a linear fit ansatz 
according to the aforementioned leading order behaviour. 
Due to the observed curvature arising from the non-leading finite volume 
corrections, only those volumes with $L_s\ge 16$ have been 
included. One also finds that the 
infinite volume extrapolation can be reliably performed 
at ranges of lattice volumes 
from $12^3\times 32$ to $24^3\times32$ if subleading corrections are considered. 

\begin{figure}[htb]
\centering
\includegraphics[width=0.40\textwidth]{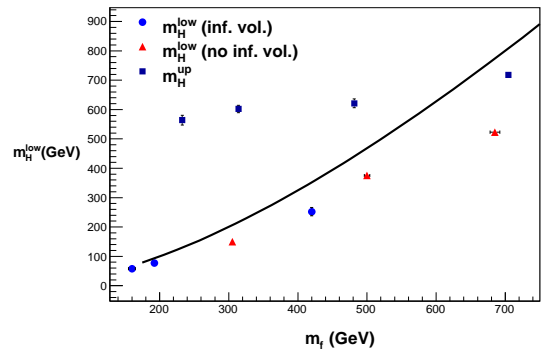}
\caption{The lower and upper 
Higgs boson mass bounds as a function of the  
fermion mass $m_{f}$ for a fixed cutoff of $\Lambda=1.5\mathrm{TeV}$. 
The solid line represents a perturbative lattice effective potential 
calculation for the lower Higgs boson mass bound. 
\label{fig:lowupmt}
}
\end{figure}

The results of the Higgs boson masses for the lower and upper Higgs 
boson mass bounds as a function of the 
quark mass at a fixed cutoff of $\Lambda=1.5\mathrm{TeV}$ 
are finally presented in \fig{fig:lowupmt}. 
All data for the upper bound have been extrapolated 
to infinite volume, as have the lower Higgs boson mass bounds at 
$m_{f} \approx 160,192,420\mathrm{GeV}$ 
Only  
lower Higgs boson mass bounds at $m_{f} \approx 305,500,685\mathrm{GeV}$ are
taken on our presently largest lattice of size $24^3\times 32$. 
However, the finite size corrections of these points will 
only provide a small change and the behaviour of the lower Higgs  
boson mass bound as a function of the fermion mass will not be significantly 
altered. 

We find that the upper Higgs boson mass bound is only moderately shifted 
by about 20\% when the fermion mass is increased from $175\mathrm{GeV}$ to $700\mathrm{GeV}$. On the contrary, the lower Higgs boson mass bound increases 
drastically with increasing quark mass. Such a significant shift of the 
lower bound has already been observed in 
\cite{Gerhold:2009ub,Gerhold:2010bh} for a quark mass of $m_{f}\approx 700\mathrm{GeV}$. 
We also show in Fig.~\ref{fig:lowupmt} a lattice perturbative 
computation of the lower Higgs boson mass bound using the effective
potential. The result of this calculation, which is discussed in the 
next section, is represented by the solid line in 
Fig.~\ref{fig:lowupmt}. Even up to fermion masses 
of $m_{f}\approx 700\mathrm{GeV}$ the lattice perturbative 
calculation qualitatively describes the simulation data rather well. This 
observation allows us to test the stability of the lower bound against 
the addition of a higher dimensional operator within the framework 
of the perturbative lattice effective potential. Moreover, direct Monte Carlo simulations 
with such a term were found to be well-described by perturbation 
theory~\cite{Gerhold:2010wy} within a small range of couplings of this higher dimensional operator.   

\section{Effective potential}
\label{sec:effectivepotential}

In addition to non-perturbative numerical data, it is instructive to calculate 
the lower Higgs boson mass bound perturbatively using the effective potential. 
To this end we follow Ref.~\cite{Gerhold:2010wy} and calculate the 
effective potential to 1-loop in the large-$N_f$ expansion.

It should be noted that these calculations are performed with the same lattice 
regularization used in the simulations 
maintaining finite spatial and temporal extents. Thus, the loop corrections are calculated numerically, by explicitly performing sums over 
the lattice momenta. An infinite volume extrapolation of these 
perturbative results is also performed.  

The vacuum expectation value and lower Higgs boson mass bound are obtained from this effective potential in the usual way, namely 
\begin{align}
\frac{d}{d\bar{\phi}} V(\bar{\phi})|_{\bar{\phi}=v} = 0
\\\nonumber
\frac{d^2}{d\bar{\phi}^2} V(\bar{\phi})|_{\bar{\phi}=v} = m_{H}^2,
\end{align}
where $\bar{\phi}$ denotes the average value of the scalar field. 
The results of such a determination are shown in Fig.~\ref{fig:lowupmt}. 
Although quantitatively different, 
the numerical data agree with the qualitative behaviour of the perturbative 
result even for rather large fermion masses. 

Based on this agreement, we estimate the effect of an additional higher-dimensional operator in the scalar field potential using the 
perturbative expansion discussed above. To this end we add to Eq.~(\ref{eq:DefYukawaCouplingTerm}) 
the term 
$\lambda_6\phi^6$. It has been demonstrated~\cite{Gerhold:2010wy} that 
perturbation theory successfully describes non-perturbative numerical results for the range $\lambda_6 = [0,0.01]$. Therefore, at this preliminary stage we examine couplings in this range only.  
Furthermore, in the presence of such a 
term we must modify the stability criterion of the effective potential.

Due to the triviality of the theory, the cutoff $\Lambda$ 
cannot be completely removed while maintaining non-zero values of the 
renormalized quartic and Yukawa couplings. 
However, the 
existence of a scaling regime suggests that predictions of the model are 
affected only mildly 
by small cutoff effects provided that the cutoff is larger than 
the relevant scales in the theory, namely $v_{R}, m_f, m_H$. It has been 
determined in the pure $\phi^4$ theory~\cite{Luscher:1987ay} that a suitable 
definition of the scaling regime is the situation where $m/\Lambda < 0.5$, where $m = v_{R}, m_H$. 
We adopt here this observation also for $m_f$, keeping $m_f/\Lambda <0.5$. 

It is in this scaling regime only that results which are universal up to small cutoff effects can be obtained from
our model. Therefore, the vacuum stability criterion should ensure that a 
stable vacuum exists throughout the scaling regime. 
Indeed, the running of the renormalized quartic and Yukawa couplings (and thus 
vacuum stability considerations) have been demonstrated to be severely 
regularization dependent outside of the scaling regime defined above~\cite{Fodor:2007fn}. 
Therefore, we choose as our stability 
criterion 
\begin{align}
\frac{d^2}{d\bar{\phi}^2}V(\bar{\phi}) > 0 , \quad \bar{\phi} < 0.5 . 
\end{align}
In the $\lambda_6=0$ case, this results in the well-known stability criterion $\hat{\lambda} \ge 0$.  
We have examined the effect of a finite $\lambda_6$ in the range 
$\lambda_6 = [0.0, 0.01]$ which results in a less than 1\% difference in the 
lower bound. However for larger values of $\lambda_6 \sim 0.1$, deviations 
as large as 15\% have been observed in the lower Higgs boson mass bound. As we are unsure of the applicability of perturbation theory in this region, further 
non-perturbative numerical simulations must be performed to properly assess the 
effect of this higher dimensional operator for larger values of $\lambda_6$. 

\section{Outlook and Conclusions}
\label{sec:conclusions}

In this letter we have performed a non-perturbative lattice 
investigation of the lower and upper Higgs boson mass bounds 
in an extension of the Standard Model with a heavy fourth quark generation. 
Since the heavy quark masses lead to large values of 
the Yukawa coupling, such a non-perturbative computation is necessary 
in order to have reliable results for the Higgs boson mass bounds at 
large $m_f$. We include only the dominant interactions, which are expected to 
be the Higgs-Yukawa interactions of the heavy fermions. In particular, we 
neglect all gauge fields.  

Our chirally invariant lattice regularization of the Higgs-Yukawa 
sector of the Standard Model (as proposed 
in Ref.~\cite{Luscher:1998pqa}) obeys a global $\SUtwoTimesUoneY$ symmetry at  
finite lattice spacing, providing a formulation that preserves the chiral nature of the Higgs-fermion couplings. 
In this setup, for a fixed cutoff of 
$\Lambda=1.5\mathrm{TeV}$, we have studied a range of quark masses  
$175\mathrm{GeV} \lesssim m_{f} \lesssim 700\mathrm{GeV}$. We find that 
the upper Higgs boson mass bound is only 
moderately increased by about 20\% for larger 
quark masses. The lower Higgs boson mass 
bound however, changes quite substantially as is summarized in Fig.~\ref{fig:lowupmt} and assumes 
for e.g. $m_{f}\approx 500\mathrm{GeV}$ a value of $m_H^\mathrm{low} \approx 500\mathrm{GeV}$. 
This puts severe constraints on the existence of a fourth generation of quarks, given the current status of direct searches~\cite{CMS,ATLAS}.  

Confronting the non-perturbatively computed lower Higgs boson mass bounds 
with a lattice perturbative calculation 
of the effective potential, we found that the effective potential describes
the simulation data rather well, even for $m_{f}\approx 700\mathrm{GeV}$. 
Based on this fact, as well as previous numerical work, we also used 
the perturbative effective potential to test the stability of the lower Higgs 
boson mass bound against addition of a higher dimensional operator for a small 
range of couplings. 
We found that the lower Higgs boson mass bound is quite insensitive 
against such an addition, although more non-perturbative numerical simulations 
are required to examine the effect of the higher dimension operator for larger couplings. Nonetheless, this preliminary work provides some confidence that the results for the lower 
bound are robust. Furthermore, although we work with a mass degenerate quark doublet in this work, the dependence of the lower bound on the mass 
splitting was found to be small at $m_f = 175\mathrm{GeV}$, 
but will investigated at larger quark masses in the near future.

\section*{Acknowledgments}
We thank George Hou, David Lin, and Ulli Wolff for fruitful discussions as well as  
M. M\"uller-Preussker for his continuous support.
We moreover acknowledge the support of the DFG through the DFG-project {\it Mu932/4-2}.
The numerical computations have been performed on the {\it HP XC4000 System}
at the {Scientific Supercomputing Center Karlsruhe} and on the
{\it SGI system HLRN-II} at the {HLRN Supercomputing Service Berlin-Hannover}.

\bibliographystyle{style.bst}
\bibliography{cited_refs}

\begin{thebibliography}{10}

\bibitem{Holdom:2009rf}
B.~Holdom {\em et~al.},
\newblock PMC Phys. {\bf A3}, 4 (2009), 0904.4698.

\bibitem{Carena:2004ha}
M.~S. Carena, A.~Megevand, M.~Quiros, and C.~E. Wagner,
\newblock Nucl.Phys. {\bf B716}, 319 (2005), hep-ph/0410352.

\bibitem{Eberhardt:2012ck}
O.~Eberhardt, A.~Lenz, A.~Menzel, U.~Nierste, and M.~Wiebusch,
\newblock (2012), 1207.0438.

\bibitem{Denner:2011vt}
A.~Denner {\em et~al.},
\newblock Eur.Phys.J. {\bf C72}, 1992 (2012), 1111.6395.

\bibitem{Chanowitz:1978uj}
M.~S. Chanowitz, M.~Furman, and I.~Hinchliffe,
\newblock Phys.Lett. {\bf B78}, 285 (1978).

\bibitem{Chanowitz:1978mv}
M.~S. Chanowitz, M.~Furman, and I.~Hinchliffe,
\newblock Nucl.Phys. {\bf B153}, 402 (1979).

\bibitem{Gerhold:2009ub}
P.~Gerhold and K.~Jansen,
\newblock JHEP {\bf 0907}, 025 (2009), 0902.4135.

\bibitem{Gerhold:2010bh}
P.~Gerhold and K.~Jansen,
\newblock JHEP {\bf 1004}, 094 (2010), 1002.4336.

\bibitem{Gerhold:2010wy}
P.~Gerhold,
\newblock (2010), 1002.2569,
\newblock * Temporary entry *.

\bibitem{Gerhold:2011mx}
P.~Gerhold, K.~Jansen, and J.~Kallarackal,
\newblock Phys.Lett. {\bf B710}, 697 (2012), 1111.4789.

\bibitem{Gerhold:2010wv}
P.~Gerhold, K.~Jansen, and J.~Kallarackal,
\newblock JHEP {\bf 1101}, 143 (2011), 1011.1648.

\bibitem{Bulava:2011jp}
J.~Bulava {\em et~al.},
\newblock PoS {\bf LATTICE2011}, 075 (2011), 1111.4544.

\bibitem{Bulava:2011ss}
J.~Bulava, P.~Gerhold, K.~Jansen, J.~Kallarackal, and A.~Nagy,
\newblock PoS {\bf LATTICE2011}, 301 (2011), 1111.2792.

\bibitem{Bulava:2012rb}
J.~Bulava {\em et~al.},
\newblock (2012), 1210.1798.

\bibitem{Luscher:1998pqa}
M.~Luscher,
\newblock Phys.Lett. {\bf B428}, 342 (1998), hep-lat/9802011.

\bibitem{Bhattacharya:2006dc}
T.~Bhattacharya, M.~R. Martin, and E.~Poppitz,
\newblock Phys.Rev. {\bf D74}, 085028 (2006), hep-lat/0605003.

\bibitem{Giedt:2007qg}
J.~Giedt and E.~Poppitz,
\newblock JHEP {\bf 0710}, 076 (2007), hep-lat/0701004.

\bibitem{Poppitz:2007tu}
E.~Poppitz and Y.~Shang,
\newblock JHEP {\bf 0708}, 081 (2007), 0706.1043.

\bibitem{Gerhold:2007yb}
P.~Gerhold and K.~Jansen,
\newblock JHEP {\bf 0709}, 041 (2007), 0705.2539.

\bibitem{Gerhold:2007gx}
P.~Gerhold and K.~Jansen,
\newblock JHEP {\bf 0710}, 001 (2007), 0707.3849.

\bibitem{Fodor:2007fn}
Z.~Fodor, K.~Holland, J.~Kuti, D.~Nogradi, and C.~Schroeder,
\newblock PoS {\bf LAT2007}, 056 (2007), 0710.3151.

\bibitem{Neuberger:1997fp}
H.~Neuberger,
\newblock Phys.Lett. {\bf B417}, 141 (1998), hep-lat/9707022.

\bibitem{Neuberger:1998wv}
H.~Neuberger,
\newblock Phys.Lett. {\bf B427}, 353 (1998), hep-lat/9801031.

\bibitem{Frezzotti:1997ym}
R.~Frezzotti and K.~Jansen,
\newblock Phys.Lett. {\bf B402}, 328 (1997), hep-lat/9702016.

\bibitem{Duane:1987de}
S.~Duane, A.~Kennedy, B.~Pendleton, and D.~Roweth,
\newblock Phys.Lett. {\bf B195}, 216 (1987).

\bibitem{Davies:1987vs}
C.~Davies {\em et~al.},
\newblock Phys.Rev. {\bf D37}, 1581 (1988).

\bibitem{Fukuda:1974ey}
R.~Fukuda and E.~Kyriakopoulos,
\newblock Nucl.Phys. {\bf B85}, 354 (1975).

\bibitem{Hasenfratz:1989ux}
A.~Hasenfratz {\em et~al.},
\newblock Z.Phys. {\bf C46}, 257 (1990).

\bibitem{Hasenfratz:1990fu}
A.~Hasenfratz {\em et~al.},
\newblock Nucl.Phys. {\bf B356}, 332 (1991).

\bibitem{Gockeler:1991ty}
M.~Gockeler and H.~Leutwyler,
\newblock Nucl.Phys. {\bf B361}, 392 (1991).

\bibitem{Luscher:1987ay}
M.~Luscher and P.~Weisz,
\newblock Nucl.Phys. {\bf B290}, 25 (1987).

\bibitem{CMS}
CMS Collaboration,
\newblock (2012).

\bibitem{ATLAS}
ATLAS Collaboration,
\newblock (2012).

\end{thebibliography}

\end{document}